# THE STUDY AND APPROACH OF SOFTWARE RE-ENGINEERING

Phuc V. Nguyen, Ph.D candidate *(Author)*
Department of Cipher & Information Technology
Ho Chi Minh City's Vietnamese Communist Party Committee
Ho Chi Minh City, Vietnam
E-mail: x201102x@gmail.com

*Abstract*— the nature of software re-engineering is to improve or transform existing software so it can be understood, controlled and reused as new software. Needs, the necessity of re-engineering software has greatly increased. The system software has become obsolete no longer used in architecture, platform they're running, stable and consistent they support the development and support needs change. Software re-engineering is vital to restore and reuse the things inherent in the existing software, put the cost of software maintenance to the lowest in the control and establish a basis for the development of software in the future.

## I. INTRODUCTION

The re-engineering re-use things already in the old software to avoid waste of material and spiritual reduce maintenance costs take to bring about the economic value, the most effective. The growth in costs and the importance of software companies, agencies, small and medium enterprises and large organizations were forced to attempt re-engineering software. Basically, the software re-engineering is to take and pass the existing software. It is the software maintenance costs, repair expensive or system architecture and performance have failed to do. We take it with technology software and hardware available. However, the difficulty here lies in understanding the current system. Usually the requirements, design and document source code (code) is not available or has expired long ago so it's not clear. What is at, functions to move? Usually the system will include features not needed; it will need to be removed in the new software system.

## II. WHEN DECIDING THE SOFTWARE RE-ENGINEERING?

When the system changes affect a subsystem and the subsystem that needs to be redesigned.
When hardware and software support has become outdated and obsolete.
When the tool supports the redesign was actually available.

## III. THE GOAL OF RE-ENGINEERING

Large number of systems built from scratch is decreasing, while the number of legacy systems in usability was very high. While the functionality of the system remains unchanged as the application environment, system level hardware and software are different. To enhance the existing functionality may be needed. Although the re-engineering efforts to improve, it does not incorporate until re-engineering is completed. This allows comparison of the functionality of existing systems and new systems.

The problem is that systems are in use today, the basic system, to be a lack of well-designed structure and organization of code changes the whole software system is difficult and expensive. Corporations do not want to destroy the system because it was built for many subsidiaries of the Group which, if destroyed will result in the application process may have made will be lost. Often the developers of the legacy systems are not always acceptable or correct the information that was lost this is the only remaining source of the code of existing software. The initial cost for developing logic and the components of the system software should not be wasted. Therefore, re-use through re-engineering is desired.

The challenge in software re-engineering is to take existing systems and imbued with good features and attributes of the software developer, created a new system where the goal is to maintain the required functions in the application of new technologies. Although the specific objectives of a re-engineering task are determined by the objectives of each company, but overall achievement has four main objectives of the re-engineering software:

• Prepare for enhanced functionality.
• Improve maintenance.
• Access to the new platform.
• Improved reliability.

Although re-engineering should not be taken to enhance the functionality of existing systems, it is frequently used to prepare for advanced functionality. The legacy system, through amendment and supplement the annual improvement in error or by more difficult or expensive. The source (code) no longer clears, and reference materials may not exist and if there is often outdated. Re-engineering determines the characteristics of existing systems can be compared with the specifications of the desired system. Reconstruct the target can be improved to facilitate the improvements. For example, if you want to improve the system based on object-oriented design, the objective of the system can be developed using object-oriented technology to prepare for the enhanced functionality to the system base. As the system developed and advanced, maintenance costs increased due to changes more difficult and time consuming. The goal of re-engineering is to redesign the system with the appropriate function module and clear interface. Document internal and outside addition will be in circulation thus improving maintainability.

Computer industry is developing at very high speeds. New hardware and system software include new features become obsolete quickly. Since the system changed so that skilled workers move to work with newer technology, leaving some employees to maintain the old system. In a short time, the manufacturer no longer supports addition. Software products and hardware become quite expensive to maintain. Even as the compatibility of legacy systems with newer ones. That's why software companies have to move to newer platform hardware, operating systems, and programming languages.

Wednesday goal is to achieve higher reliability. Although reliability cannot always be very high, more likely, over time and may have changed on the reliability, the system time much has changed, and it raises many problems such as maintenance and continues to change. For maintenance and further changes to the software reliability will decrease until no longer acceptable.

## IV. DEFINITION OF SOFTWARE RE-ENGINEERING

1. GENERAL MODEL OF SOFTWARE RE-ENGINEERING

Re-engineering starts with the source code of the basic system exist, and ends with the source code of the system will target. This process can be as simple as using the translation tool to translate source code from one language into another language (FORTRAN to C) or from one operating system to other operating systems (UNIX to DOS). On the other hand, the task of re-engineering can be very complex, using existing source code to recreate the design, determine the requirements in existing systems and then compare them to current requirements now, remove things no longer apply, the system design using object-oriented design and the last code into a new system. Figure 1 the general pattern of re-engineering software for engineering and process for all levels of re-engineering based on the level of software development level abstractions mentioned above.

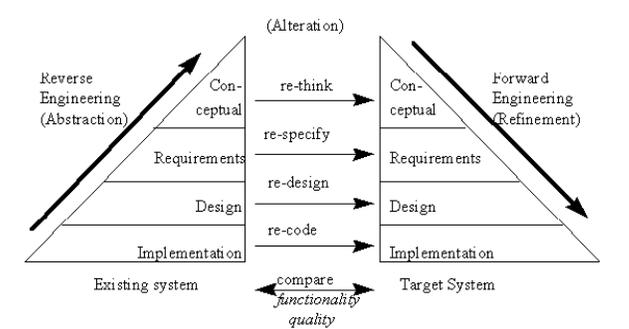

**Figure 1 the general pattern of re-engineering software**

The model in Figure 1 applies three principles of re-engineering: abstraction, the amendment and refinement. This abstraction level is a gradual increase in the level of abstraction of the system. The present system was created by replacing a row of information is the information that is more abstract. Abstractions make the description emphasize the characteristics of the system. The upward movement is called reverse engineering and related accessories to the process, tools and techniques. The amendment is to create one or more to convert a representation of the system without changing the level of abstraction in which additional, delete and modify information. The refinement (Refinement) is the gradual reduction in the level of abstraction of the system caused by the continuous replacement information in existing systems with more detailed information. It is technically forward (Forward engineering) as software developers with the code (code) but with some new screening process.

To change a feature of the system, the work is done at the level of abstraction at which information about characteristics that are clearly presented. To translate the code (code) is a language aiming to reverse engineering are necessary, changes made at the level of implementation.

2. REVERSE ENGINEERING

Reverse engineering is a process of analysis to determine the relationship of the system and create the components of the system in another form or in a higher level of abstraction. In reverse engineering, and design requirements necessary structure and content of the system must be decreased. In addition to retaining the schema relations and interactions, information and rules on business applications and processes useful in the business must be saved. Key objective of creating industry back to replace, restore the lost information, detecting side effects, and facilitate the reuse efficiency of the process affect the success of the project re-engineering. Reverse engineering does not change the system created a new system; it is an inspection process does not change the overall functionality of the system.

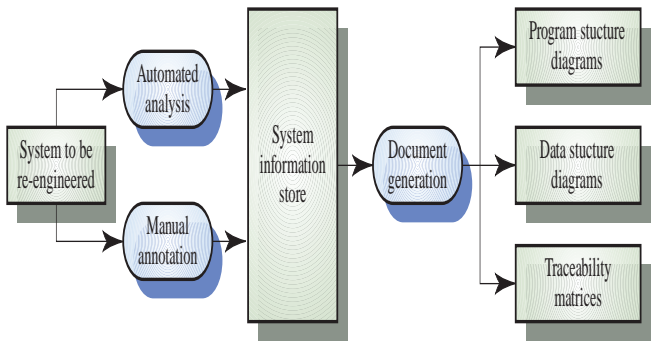

**Figure 2. The process of reverse engineering**

Reverse engineering often precedes re-engineering. Sometimes the reverse engineering is preferred. When the specification and design of the system must be determined before using them as input for the specification of requirements for replacement systems. When the design and specification for a system are needed to support the operation and maintenance program.

3. FORWARD ENGINEERING

Target system is created by moving downward through the level of abstraction (the levels of abstraction). A decrease in the level of abstraction of the system is described by replacing a row information system exists for much more detailed information. The movement to actually move through the process of software development standards so that industry forward. Technology moves forward from high-level abstractions and logical implementation independent design of the physical features of the system. A series of requirements through design to implementation is followed. These risks are due to technical level and in the preparation of reverse engineering. The project is exposed to more risk with the change or addition of new requirements. So the industry forward:
• Sometimes referred to as rehabilitation or renovation.
• Restore the design information from existing source code.
• Use this information designed to reorganize the existing system to improve the overall quality or performance of it.

## V. PROCESS RE-ENGINEERING

• Inventory analysis:
+ Classification of software application's critical work by age, current maintenance, and other criteria.
+ Determine the candidate process re-engineering the software.

• Select document reorganization:
+ Maintain old documents.
+ Updated the old documents if they are used.
+ Rewrote the complete document for important system to focus on the minimum necessary

• Industry is back:
+ Restoration design.

+ Anatomy of a program to create a representation of the program at some level of abstraction higher than the source.

• Reorganize code (Code restructuring):
+ Source code analysis and violation of structured programming practices are recognized and corrected.
+ Modify the code should be reviewed and tested.

• Reorganization of data (Data restructuring):
+ Usually requires reverse engineering altogether.
+ Architecture data currently fragmented.
+ The data model is defined.
+ Data's structure is being reviewed for quality.

• Technology transfer.

## VI. THE APPROACH RE-ENGINERRING SOFTWARE

There are three different methods to access the software re-engineering. The approaches differ in quantity and rate of replacement of existing systems to the target system. Each approach has benefits and risks of its own.

1. Approach BIG BANG
BIG BANG approach to replace the entire system at a time, as shown in Figure 6, this method is often used for projects that issue to be resolved immediately. Such as moving to a system with different architectures.

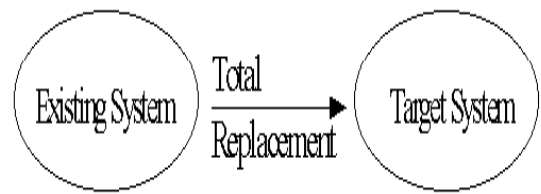

**Figure 3. Access to BIG BANG**

The advantages of this approach are the system put into a new environment all at the same time, there are no old and new interfaces between components must be developed. The downside is the result of the project may not always be appropriate. For most large systems, this approach consumes too many resources or requires large amounts of time before generating the desired system. The risk of this method is the very high, one challenge is that the change of control; time between new systems and complete systems have changed and will likely be made to the old system and must be reflected in the old system. That is what is being re-engineering is likely to change.

2. Added Method (Incremental Approach)
Approach, "added" to re-engineering is also known as "Phase-out". In this approach, as shown in Figure 4, the part of the system is re-engineering and incremental updates as a new version of the system needed to meet new goals. The project is divided into the re-engineering based on the portion of the existing system.

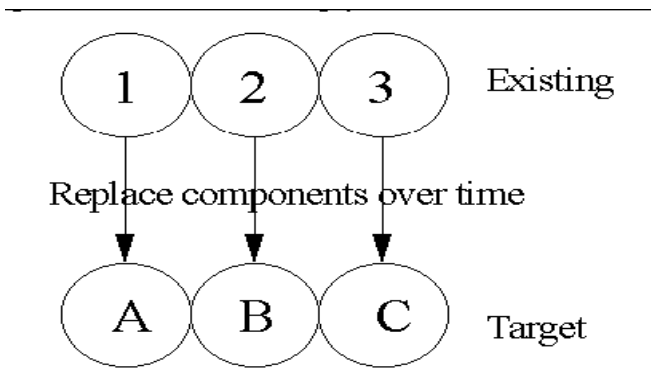

**Figure 4 Approaches to increase (Incremental)**

The advantage of this approach is that components of the system are produced faster, and it is easy to control errors when new components are clearly defined. Since the interim version was released, the customer can see progress and quickly identify the function is lost. This method changes the old system of easy-easier than changing the system has no impact on existing components. The downside is that the system takes longer to complete with a temporary version that requires carefully controlled. The whole structure of the system cannot be changed, only the structure of the specific components is recycled. This requires careful in identifying the components in existing systems and planning structures for the target system. This approach has lower risk compared with the BIG BANG because each component of the re-start the industry. The risk of code can be identified and tracked.

3. The method of evolution (Evolutionary approach)
This method is similar methods rise. Part of the original system is replaced by the newly re-designed system. However, in this method, the parts are selected based on their function, not based on the existing system structure. Target system is built using the functionality of the necessary adhesion. This method allows developers to focus on re-engineering to identify the object function in the current system. As shown in Figure 5, the components of the current system are divided by function and re-engineering the new component.

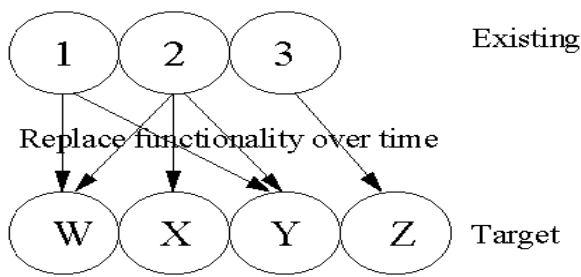

**Figure 5. Approaches to increase (Evolutionary Approach Re-engineering)**

The advantage is that the resulting modular design and reduces the scope for a single component. This method works well when converted to object-oriented technology. The downside is that same function must be defined in the current system and then refined as a single functional unit.

### VII. THE STAGE AND TASK OF RE-ENGINERRING SOFTWARE

This is the core processes that every organization should follow when re-engineering. Re-engineering poses technical challenges of its own and without a comprehensive development process will waste time and money.
Automation and tools can only support this process. Re-engineering can be divided into five stages, and the work involved, starting from the initial stage of determining the feasibility, effectiveness and cost of re-engineering and finish the transition to the target system. In that period are:

• Establish re-engineering.
• Analyze the feasibility of the project.
• Analysis and planning.
• Implement re-engineering.
• Conversion and testing.

1. Establish Re-engineering
This group is the project manager software re-engineering effort from the start to finish the project and will need comprehensive training how to manage the technological change, the basics of re-engineering technology and the use of development objectives and the maintenance process and maintenance. Their mission will start with a very diverse set of objectives, strategies and action plans in the current environment based on identified business needs, including repair costs. Although the team members have skills in software development standards, they will need more specific skills. They will be responsible for identifying; testing and purchasing new tools, then make sure employees are properly trained on the tool will use them effectively. The group will need to provide an internal marketing re-engineering work, consulting with staff to verify the process is being applied is correct. This task requires the team members must have good communication skills to resolve not to accept new concepts and perception of software ownership. When the field of re-engineering progresses, the team members will need to continue studying the technology.

2. Analyze the feasibility of the project
The initial work of the re-engineering team was to assess the needs of the organization and goal is that systems need to respond. It is important to re-engineering strategies are consistent with the standards of the organization or not? Software products currently in use need to be analyzed in terms of issues of specification, including goals, motivations, constraints and regulations of the work. The value of the application must be investigated to determine what has been achieved from the efforts to re-engineering: the level of software quality is expected to increase, improve the maintenance efficiency and value work is improved. Once it reached the expectations, they must be expressed in a measurable way - reduce maintenance costs, improve performance ... then the cost of re-engineering should be compared with the desired information cost and value of the system.

3. Analysis and planning

Re-engineering the software needs to do in three steps: analysis of existing systems, identify the characteristics, features of the target system, and create a standard set of tests to confirm proper transfer function Affairs. The analysis step begins by locating all the available code and documentation, including user manuals, design documents and specifications required. Once all the old system information is collected, it is analyzed to determine the direction only. Current status of the current system, performance and maintenance must be specified to determine the capital for the target system. A set of software metrics have been selected to assist in the identification of quality problems with the current system and the priority of applications those are candidates for re-engineering the technical quality of medical and value of work. The measurements should include the cost of software change, if increased maintenance is one of the objectives of the process re-engineering. At the end of the process re-engineering, metrics should be used to determine the quality of the new system. The collection of data on the new system should continue to grow throughout to know if what is happening to be normal or expected, and to respond quickly to signs of abnormalities. After the current system and its quality characteristics have been specified, the steps clearly desired characteristics and specifications of the target system start. The characteristics of the system are to be changed are specified, such as operating systems, hardware platforms, structural design, and language. Finally, a standard testing and certification must be created. This will be used to demonstrate the new system is functionally equivalent to the old system and the function remains unchanged after re-engineering. The test for the target system can be implemented to gradually increase if necessary, but can map out the functions of the base system is important.

4. Implement re-engineering

Now the goal of re-engineering has been clearly indicated. The approach has been identified, and the base system has been analyzed, reverse engineering and forward is starting. Using abstract level in Figure 2, the actual function of the system is off base launched by reverse engineering. The various tools available for this task. These tools must be reviewed to see if the usability in the context of the objectives of the process re-engineering. We must integrate easily into the process.

After the desired level of abstraction has been achieved, the technical transition can begin. Technology transition to match properly with the standard procedures followed derivative software development. That is, if the software was redesigned to fit an overall architecture of the new system, the process of reverse engineering to extract the software requirements and process industries will begin to transition to the development of a new design. In the industry forward, any change or increase in function must be avoided as it complicates the process of certifying the validity. During these stage, quality assurance and management disciplines and techniques to be applied. Measurement techniques should continue to use to evaluate the improvement of software and identify potential risks.

5. Conversion and testing

As the functionality of the system development, testing must be performed to detect errors in the process re-engineering. Experimental techniques and methods that work began to develop the system. Assuming that the requirements for the new system inherited from the system of testing and test development in the planning stage can be used. The same test cases can be applied to both legacy systems and future systems, comparing the results to confirm the functionality of the system. Documentation for the succession to be updated, rewritten or replaced at this stage to apply the new system, and contains the information necessary to operate and maintain it.

VIII. THE RISK OF RE-ENGINERRING SOFTWARE

SATC has coined the phrase "Hybrid Re-engineering" means the process re-engineering not only use but also a combination of different levels of abstraction and methodology changes to forward a system that has with a target system. The Hybrid Re-engineering projects choose to combine the successor based on the conditions of the old system, the needs of the project and the budget and plan. In the Hybrid Re-engineering, system's re-engineering have been using the approach shown in Figure 6, a process of adaptation of the general model for software re-engineering the application display.

In Figure 6, only three are used development. The first is just a process of translation from existing code into a new language. Monday, tracking the use of existing code to identify the request can be accepted by the application of COTS packages. Tuesday there are many process's re-engineering standards. The development of new code required for the project cannot be satisfied by one of the other, and it is attached together to move and combine with the COTS parts.

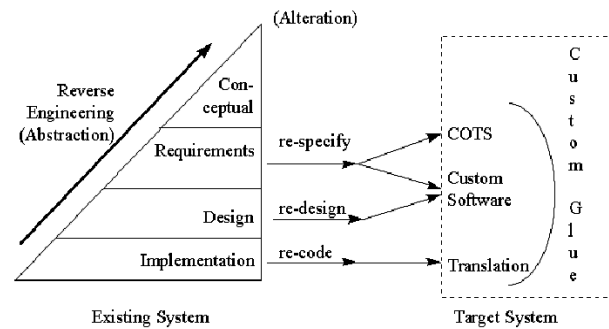

**Figure 6. Track Hybrid Re-engineering**

Re-engineering as a methodology for the development of the capital risks, such as schedule, functionality, cost, and quality. Hybrid Re-engineering has been developed to reduce some of the risks of COTS packages can be expected to have high reliability and require minimal development time. A method of reducing time and costs through the Hybrid Re-engineering while maintaining functionality through a process of translating directly from the existing code to new languages. Hybrid re-engineering is an improvement; try to combine three different Re-engineering, so the risks normally associated with re-engineering can be increased by combining the risks to each direction (track). When the hybrid re-engineering are combined with products from many directions (track) developed (COTS, custom software and software services), a new risk is the interface and interoperability of products. For example, transfer data between products may cause the possibility of compatibility issues and time; the COTS package may not work as expected.

In the combined data can be used by managers to improve the software-development effort and reduce the risk. The data can indicate how a project meets its objectives. In hybrid re-engineering, data can support the decision on follow-up options for different software components and functions.

1. Track Hybrid Re-engineering

The following sections describe each of three directions (track) hybrid re-engineering. After each track is described, the risks related to the track are determined and the data is determined accordingly.

a) Translation Track Hybrid Re-engineering

Figure 7 is a diagram of a typical software system has been used for some time. In the re-engineering example, assume that the project is converted from FORTRAN C + + (but not necessarily toward an object-oriented design).

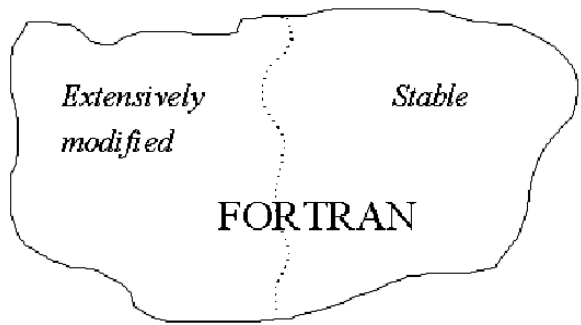

**Figure 7 Current software system**

Looking at the system software current system, we can see two different classifications of code (code), stable code that has transformed the minimum, and keep those requests, and some code has many changes and has become unstable, unreliable, and pays dearly to maintain. Re-engineering a stable code may not require all reverse engineering (reverse ), It it can perform a simple division of rewriting the code in the new language or new environments. This process is considered Translation Track Hybrid Re-engineering. This process is shown in Figure 6, the code in the current system is relatively stable, have minimal changes to the original design and architecture, need to be determined. This can be done by analyzing the code and change the report. Before re-engineering, in determining the candidates to offer, such as the complex logic and data calls to provide valuable information about the structure and matching of candidates and proposals for service. Which defines the components have been widely maintained. Change or problem reports provide data. Subscribe to the criticality function will aid in the decision.

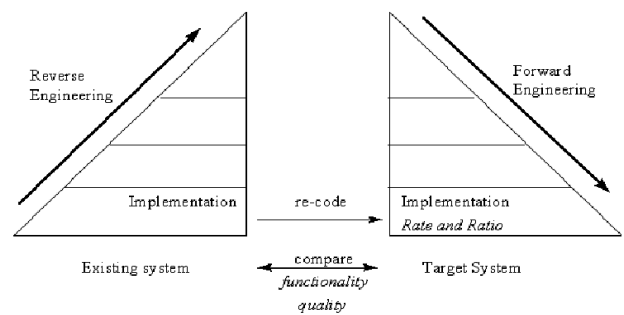

**Figure 8 Translation Track Hybrid Re-engineering**

In this process, the main risk is the quality of the resulting code. When switching from one language to another language, the code may have the syntax of programming languages, but no new structures or new features. Many source code for supporting the transition from one language, any operating system into the language, other operating systems. The source code cannot resolve service problems. The translation line by line does not take advantage of new language, structure, etc., usually as the result code is "C-TRAN" - syntax C on the structure of FORTRAN. Whereas the old code is standard, this does not ensure that the new code will have similar quality. If you do not achieve the quality, the code needs to be improved. If 20-30% service code to be changed to improve its quality or to meet the standards, the code should not be used and all the functions and composition of re-engineering should be the other way. Identify the functions of the base system will provide a basis to estimate the complete new system of how during the development process and provide an assessment as to complete the system. One method being tested by the SATC and any other companies to follow the evolution of industrial re-use function points as a measure of the function. In this application, using the method of counting the basic functions described by Albrecht, an approximate estimate can be collected as the type and total count of tasks. This can be used as a starting point in promoting the comparison of the transfer functions between the base system and target system. Tools available to scoring functions from COBOL code and SATC are working to develop a tool for FORTRAN and C. In assessing the evolution of translation. One of the measures can be evaluated at which the function. Approximation by the functions is moved from the old system to the new system. Once complete re-engineering, it is important to verify the functionality is retained in the new system, as well as the quality of the code has been improved. There is no simple method to ensure the function was transferred to the system. The most common method is the use case involves running tests on the original system and then repeats the test on a complete new system. I Furthermore, need to ensure that relevant documents have been updated.

b) COTS Track Hybrid Re-engineering

In COTS Track Hybrid Re-engineering, the requirements and functionality have been identified can be implemented using COTS.

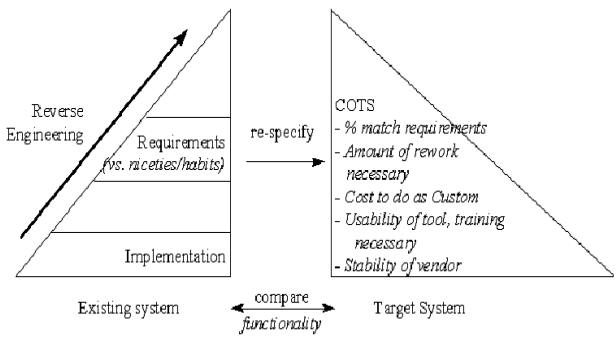

**Figure 9 COTS Track Hybrid Re-engineering**

After applying the techniques of reverse engineering software to define the requirements, it is important to separate requirements that must be contained in the target system from the requirement that users want in the new system, because it has become a habit or comfort with the features. By using the advantages of COST is to reduce development time and improve reliability. Testing is still needed in environmental harmony. The disadvantage may be a number of requirements to be satisfied by the package. For example, if the system is a specific field is 10 characters long and only allowed eight COTS package, whether this is acceptable? Although the use of COTS software development to reduce the time and increase reliability, COTS also introduce more risk. A major risk is that the software will not be advertised or as expected, that it is unreliable, immature or incomplete. The software can also be frequently asked manufacturers constantly improved version upgrade. In worse cases, this change can change or remove the essential functions of the system. COTS may require amendments or additions to fit the requirements, increase the duration or reducing reliability.

In addition, the use of COTS may limit the further improvements to the system, because changes in the COTS provide functions cannot occur because the result is acceptable.

The stability of the provider should also be part of the evaluation process because it may be necessary for them to change the request later. Another added cost can be do not familiar with COTS. These simple changes can be used, as the new icon, which will require additional training time. Some measurements are used to help reduce the risk associated with the choice of the COTS package. It is important first to identify the percentage of the requirements are demanding that the appropriate package and request that the package is not entirely appropriate. This information can be used to determine how many work packages or additional packages required meeting the system requirements. Improvements or additions will affect the duration and budget, and may affect retention and reliability. After all this evaluation is completed, the cost to develop from the start should also be evaluated, including time tested, and compared to the total cost of COTS.

In this case, COTS are executed, in terms of functionality of existing systems must be reconciled with the function of the target system. The process of comparison must be based on testing, as is done for the function code interpreter.

c) Custom Track Hybrid Re-engineering

Custom Track Hybrid Re-engineering is shown in Figure 10; similar to traditional re-engineering since the new code is transferred from the old system exists.

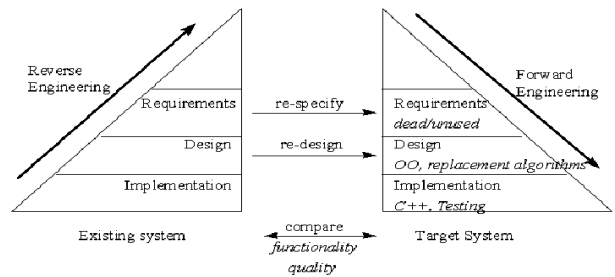

**Figure 10 Custom Track Hybrid Re-engineering**

Here, reverse engineering is done first. The function is not satisfied by the COTS package or by the translated code must be chemical homogeneity, and their requirements and plans are summarized. Technology transfer is done at that time. Starting with requirements analysis, with the aim of identifying the requirements are not necessary. The process then is similar to any development process, starting with the development of a new design, with object-oriented structure, if required, then execute the code and testing. The advantage that the resulting code must meet the exact requirements. The code was developed to high quality and well structured, requiring little maintenance repair. Difficulties similar to software development standards, in which the code cannot be trustworthy, need more testing, and that the development / testing process can overcome time and budget costs.

Custom code as well as risks due to any code, from the quality, reliability and duration. Since most of the functionality of the old system was identified as unique to the system will be implemented with custom code, the risk is one of the features will not be realized and the functionality of the system will lack. The characteristics identified as critical to the system which is complete with custom code will require extensive testing. The measure for this process is a combination of both the measurement process and product measurements. Before reverse engineering, the quality of the existing system should be evaluated for comparison later with the target system. Results obtained in the course should be saved tracks to help evaluate the cost of re-engineering. This can also be used to approximate time of completion and use of evaluation 60% of the time is re-engineering. A request is specified. The quality of these can be evaluated to determine not testable. This was also mentioned before the main function used to calculate the conversion rate in terms of functionality with the target system. Code analysis tools can be used to assess the quality of the code as it is developed and identify risk-free. In testing, inconsistency or error rate helps to assess reliability. In this case, both the function and quality were compared the existing systems with the target system.

2. Hybrid Re-engineering approach

Hybrid Re-engineering requires a re-engineering approach to tradition, but with the additional need. When you start re-engineering, the initial argument for re-engineering such as cost and quality are developed and expectations, such as return on investment have been provided. Systems's analysis is the basis to do to determine the feasibility of the Hybrid Re-engineering. Analysis system should provide the basis of principles in determining the optimal strategy and the cost for the project. Once you decide to use the Hybrid Re-engineering, additional analysis is needed.

The first step in the Hybrid Re-engineering approach is to investigate the request and that the force of the development. These factors include setting a timetable for reverse engineering and forward. Time must be built to test to investigate available COTS. In the process of the industrial development transition period should decrease with the use of COTS and code translation, more time will be required to test relevance and interface products. Budget constraint must also be considered. Trust management and the needs of the organization must also be determined. Three steps of development, balance is needed so it is important to prioritize requirements.

The next step is to analyze deeply the legacy system, focusing on functions and the appropriate code for each step (Translation, Custom, COTS). The software re-engineering, analysis of the current system is usually done to provide quality evaluation of existing systems and maintenance costs. This information is used to justify the cost and improve at the end of the project. While these reasons are still relevant in the Hybrid Re-engineering, additional features of the legacy system must now be studied carefully. In the analysis of legacy systems, components and functions must be defined. They must be further evaluated to determine the materials available determine the features you need from what is no longer needed or what the user has become accustomed. The source must be scheduled by maintenance costs, and quality of the existing structure. Unique function for this project must be identified. Once the code is divided into tracks, each track will be conducted independently. The schedule for completion of the track will vary based on the task. As the track consists of various tasks, the unity of the final product can begin. After the system is complete and all the best track, there are two tasks: to try and prove it. Firstly, comprehensive and integrated system testing must be done to ensure all components work together as a unit and to ensure all the functionality of the system has been transferred the new system. Monday, justify the re-engineering is required - do the benefits justify the costs. Some expected benefits such as improved maintenance and operating costs, can only be demonstrated indirectly through the improvement of quality. Improving the quality of the original can be evidenced by analyzing data from the old system compared to the new system. Because the new system was put into operation, additional data can be used to verify the improvements.

3. Risks of Hybrid Re-engineering

All software developments have inherent risks of the schedule and cost. Hybrid Re-engineering as a methodology for software development. Because of its components in three-track development of diversity, is subject to the risks discussed in each track description. In addition, the Hybrid Re-engineering as a method of re-engineering software risks only added functionality and quality; the functionality of the system is to be preserved in the new system, and quality improvement, implying lower operating costs and maintenance.

4. Benefits of Hybrid Re-engineering

Overall, Hybrid Re-engineering is done against the construction of a new system, because the pattern of job application procedures, and logic built into the software. The process can be deeply embedded in the procedures as simple as a field or the length of the complex as an algorithm; the only source of information in the code to the old system. Monday demonstrated a re-engineering is the development and maintenance costs of older systems; time for developing logic and components are not wasted. In re-engineering, system is being re-implemented with software development methodologies well, properties, and new technologies while maintaining the current function. Reliability and maintenance are also improved.

Hybrid Re-engineering has the additional benefit of a reduced development schedule, thus reducing costs. Development schedule is shortened first by reducing the number of reverse engineering. Using the minimum working time of reverse engineering. The use of COTS reduces the development industry forward and time-tested and therefore, lower costs. Selecting the right COTS also increases reliability when these packages have been tested extensively.

## IX. THE RISK OF RE-ENGINERRING SOFTWARE

Although re-engineering is often used as a means to reduce the risk, reduce costs and maintain the operation of deriving the software, but re-engineering cannot avoid risk. The risk of early identification programs to support and project management in the preparation of estimates, risk assessment of software re-engineering and provide a practical framework for the expectations. Identifying risks is essential for effective risk assessment, risk analysis and risk management.

## X. CONCLUSION

As the software industry moves to a new millennium, many software designers in developing new methods, improved software and reduce development time and maintenance. However, most companies have legacy systems that are overdue and save costs to maintain. The system cannot only be replaced by the new system, which contains company information and implied that would be lost. They are an investment, and too expensive to develop and evolve just to remove. For re-engineering to become a useful tool to convert old, outdated systems to the more efficient, logical system. However, development projects often short of time and money, making the need to look at the replacement necessary. The use of COST package is considered a way to increase reliability while reducing development and testing time. Translation as a means of reducing the time and cost. This results in a combination of measures to develop a form of re-engineering. Re-engineering is a structured in software development and different approaches and specific stages and tasks. However, extensive work is still needed. Although it is noted that the system administrator is very important, no method for incorporating them. There is not a method for evaluating the quality of the new COTS system. Data are also needed during the re-engineering, but it is not clear exactly how the concept can be considered and expected that the actual data, evaluate what they are proposing to define quantify. Paper is a summary of the strategies and techniques in re-engineering, to serve as a basis for future work.